\documentclass[pra,aps,twocolumn,showpacs,tightenlines,amsmath]{revtex4}
\usepackage{graphics,bm}
\usepackage{graphicx}

\begin{document}

\title{Bogoliubov theory of Feshbach molecules in the BEC-BCS crossover}
\author{M. W. J. Romans,$^1$ and H. T. C. Stoof$^1$}
\affiliation{\it $^1$Institute for Theoretical Physics, University of Utrecht,
                 Leuvenlaan 4, 3584 CE Utrecht, The Netherlands.}

\date{\today}

\begin{abstract}
We present the Bogoliubov theory for the Bose-Einstein
condensation of Feshbach molecules in a balanced Fermi mixture.
Because the Bogoliubov theory includes (Gaussian) fluctuations, we
can in this manner accurately incorporate both the two-body and
many-body aspects of the BEC-BCS crossover that occurs near a
Feshbach resonance. We apply the theory in particular to the very
broad Feshbach resonance in atomic $^6$Li at a magnetic field of
$B_0 = 834$ G and find good agreement with experiments in that
case. The BEC-BCS crossover for more narrow Feshbach resonances is
also discussed.
\end{abstract}


\maketitle

\section{Introduction}
Since the achievement of Bose-Einstein condensation (BEC) in an
atomic Bose gas \cite{Anderson95,Hulet95,Davis95}, the field of
ultracold gases has been one of the most active fields in physics.
This is primarily due to the fact that these gases are very clean,
accessible, and easy to manipulate. For both theorists and
experimentalists, ultracold atomic gases have therefore become an
important playground for the study of a diverse set of quantum
phenomena, that often in other guises also appear in very
different condensed-matter systems. At present also degenerate
atomic Fermi gases are available \cite{Jin} and have been at the
center of attention in the last two years. In particular, the
superfluid Bardeen-Cooper-Schrieffer (BCS) phase has been created
in two-component Fermi mixtures
\cite{regal2004,MIT,Duke,Innsbruck,ENS,Rice}. To overcome the
problem of the small critical temperature, which depends in
general exponentially on the interaction strength, all experiments
have made use of a Feshbach resonance
\cite{Feshbach62,Stwalley,Tiesinga}. In this manner the
interactions between the atoms can be strongly enhanced by an
external magnetic bias field, giving rise to the BEC-BCS crossover
phenomenon \cite{Eagles,Legget,Nozieres}. As a result of the
atomic physics of the Feshbach resonance, the nature of the Cooper
pairs in the BEC-BCS crossover is, however, not solely determined
by the interaction strength or scattering length, but in principle
also depends on the width of the Feshbach resonance. In the limit
of an infinitely broad resonance, the properties of the gas can be
derived from a single-channel theory that requires only the
resonant scattering length as an experimental input. In general,
however, a two-channel theory is needed. This is in particular
true for the description of the wave function of the Cooper pairs
that plays an important role in the BEC-BCS crossover as we will
show in this paper.

In the crossover experiments with atomic $^6$Li, on which we focus
mostly in this paper, the atomic gas is spin polarized with
respect to the electron spin of the atoms. In the event of a
collision, the two atoms therefore approach each other in a
triplet state, which is called the open channel. Due to the
hyperfine interaction, there is a nonzero coupling to a
closed singlet channel, which has a different interatomic
interaction potential that, most importantly for our purposes,
contains a bound molecular state. This bound molecular state we call the bare
molecule from now on. Moreover, the triplet state has a magnetic
moment, and the energy of the two incoming atoms can thus be tuned
by a magnetic field due to the Zeeman effect. Near a Feshbach
resonance, the threshold energy of the open channel is exactly
tuned near a bound state of the closed channel and the nonzero
coupling between the two channels causes the scattering cross
section between the atoms to be resonantly enhanced, in a similar
manner as in many collision experiments discovering new elementary
particles in high-energy physics.

In this paper we aim at presenting a theory of the BEC-BCS
crossover that includes (Gaussian) fluctuations on top of the
mean-field theory. Moreover, we include the above mentioned
two-channel physics of the Feshbach resonance exactly and in this
way improve upon the various theories
\cite{falco2004a,strinati,levin,mackie,ho,griffin,burnett,wetterich,
      stringari,levin2,javanainen}
that have also been used to discuss different aspects of the
BEC-BCS crossover observed experimentally. Some consequences of
our theory were previously obtained in Ref. \cite{romans}. Here we
considerably expand upon that work by giving a detailed derivation
of our Bogoliubov, or random-phase approximation (RPA), theory and
present new results in the BEC, unitarity, and BCS limits of the
theory. The main physical idea behind the theory is that the
entire BEC-BCS crossover can be seen as a Bose-Einstein
condensation of pairs or dressed molecules, that in the extreme
BEC limit are the bare molecules, and in the BCS limit the atomic
Cooper pairs. The wave function of these dressed molecules is
given by the linear superposition
\begin{eqnarray}
\label{defz}
\langle {\bf r}|\chi_{\rm dressed}\rangle = \sqrt{Z}\chi_{\rm
m}({\bf r}) |{\rm closed} \rangle + \sqrt{1-Z}\chi_{\rm aa}({\bf
r}) |{\rm open}\rangle,
\end{eqnarray}
where ${\bf r}$ is the interatomic distance. The parameter $Z$ is
the probability of the dressed molecule to be in the bare
molecular state in the closed-channel of the Feshbach resonance.
The spatial part of the bare molecular wave function is denoted by
$\chi_{\rm m}({\bf r})$. The spin part of the bare molecule is
equal to $|{\rm closed}\rangle$, in agreement with the fact that
it is a bound state in the closed channel. The open channel has
the spin state $|{\rm open}\rangle$ and the spatial part of the
wave function in this channel is $\chi_{\rm aa}({\bf r})$. The
latter corresponds to the well-known wave function of the Cooper
pairs in BCS theory. That this is a sensible way to describe the
system at low temperatures, follows from the fact that it contains
the correct physics in both extremes. In the extreme BEC limit the
dressed molecules are just bare molecules, i.e., $Z \simeq 1$. In
the BCS limit we have $Z \simeq 0$ and the dressed molecules
correspond to the BCS Cooper pairs with the wave function
$\chi_{\rm aa}({\bf r})$. In between these two limits the
probability $Z$ changes gradually and in this manner describes the
smooth crossover from the BEC to the BCS limit. This important
microscopic parameter has indeed been measured recently for $^6$Li
by Partridge {\it et al}.\ \cite{Rice} and was shortly thereafter
calculated theoretically for that case in Ref. \cite{romans}.

The paper is organized as follows. Because of its importance for
the many-body theory that is of most interest to us, we first
briefly discuss in Sec.~II the atomic physics that is required to
account for the resonant interactions between two atoms. In
particular, we give a simple two-body derivation of the
probability $Z$, that is then generalized in Sec.~III when we
discuss in detail the many-body Bogoliubov theory for the
Bose-Einstein condensation of the Feshbach molecules. In Sec.~IV
we apply the theory to the broad Feshbach resonance of $^6$Li at a
magnetic field of $B_0 = 834$ G. We here present results for the
probability $Z$, the spectral function of the molecules, and the
speed of sound in the gas throughout the BEC-BCS crossover. In
Sec.~V we briefly discuss the qualitative difference that would occur in
these quantities if the Feshbach resonance in $^6$Li would be much
narrower. In Sec.~VI we consider the
BEC, unitarity, and BCS limits of the theory. We
end our paper in Sec.~VII with some discussion and conclusions.

\section{Two-body physics and background scattering}
For several reasons it is most convenient to start with the
discussion of the relevant two-body physics of the Feshbach
resonance. First, it is a natural way to introduce the probability
$Z$, that describes the exact bound state wave function of the
two-body Hamiltonian. Second, it gives a quantitative description
for the problem that works quite well in the BEC-limit, where
there are no free atoms in the gas and the Bose-Einstein
condensate of dressed molecules is well described by two-body
physics. Third, it illustrates how to incorporate not only the
resonant scattering, but also the background scattering of the
atoms, which turns out to be very important in the case of the
extremely broad Feshbach resonance in $^6$Li of interest to us.

For two atoms, we can simplify the Feshbach problem by first
splitting off the center-of-mass motion. In that manner we arrive
at the time-independent two-channel Schr\"odinger equation
\begin{align}
\label{twobody-schrodinger}
\begin{pmatrix}
H_{\rm aa} & V_{\rm am}\\
V_{\rm am} & \delta_B
\end{pmatrix}
\begin{pmatrix}
|\psi_{\rm a}\rangle\\
|\psi_{\rm m}\rangle\\
\end{pmatrix}
=E
\begin{pmatrix}
|\psi_{\rm a}\rangle\\
|\psi_{\rm m}\rangle
\end{pmatrix},
\end{align}
where $H_{\rm aa}=-\hbar^2{\mbox{\boldmath $\nabla$}}^2/m+V_{\rm
bg}({\bf r})$ is the Hamiltonian of the relative motion in the
presence of the background interaction $V_{\rm bg}({\bf r})$,
$|\psi_{\rm a}\rangle$ is the component of the two-body wave
function in the open (triplet) channel, and $|\psi_{\rm m}\rangle$
is the component of the two-body wave function in the closed
(singlet) channel. The bare detuning from resonance is given by
$\delta_B$, and $V_{\rm am}$ is the coupling between the open and
closed channels. We formally eliminate $|\psi_{\rm a}\rangle$ from
Eq.~(\ref{twobody-schrodinger}), by making use of the fact that
the energy of the exact molecular bound state that we want to
determine never lies in the spectrum of the atomic Hamiltonian
$H_{\rm aa}$, and insert a complete set of states in the resulting
expression. In the case of a large and positive background
scattering length, there exists a bound state $|\psi_{\rm
b}\rangle$ just below the threshold of the scattering continuum of
the atomic Hamiltonian $H_{\rm aa}$. Since this bound state may be
relevant we include it explicitly as well. In this manner we
ultimately find
\begin{align}
\label{2bodyder}
E-\delta_B&
=\langle\psi_{\rm m}|V_{\rm am}\frac{1}{E-H_{\rm aa}}V_{\rm am}
                                                 |\psi_{\rm m}\rangle\nonumber\\
\hspace{0.2cm}=&\int \frac{d^3{\bf k}}{(2\pi)^3}
\langle\psi_{\rm m}|V_{\rm am}|\psi_{\bf k}^{(+)}\rangle\frac{1}{E-2\epsilon_{\bf k}}
\langle\psi_{\bf k}^{(+)}|V_{\rm am}|\psi_{\rm m}\rangle\nonumber\\
&+\langle\psi_{\rm m}|V_{\rm am}|\psi_{\rm
b}\rangle\frac{1}{E-E_{\rm b}}\langle\psi_{\rm b}|V_{\rm
am}|\psi_{\rm m}\rangle,
\end{align}
where $2\epsilon_{\bf k}=\hbar^2{\bf k}^2/m$ are the energies of
the atomic scattering states $|\psi_{\bf k}^{(+)}\rangle$. We
focus on the first term in the right-hand side of
Eq.~(\ref{2bodyder}) that is due to the continuum of scattering
states, and leave the second term due to the bound state for
later.

We would like to use the pseudopotential
approximation for the matrix element $\langle\psi_{\rm m}|V_{\rm
am}|\psi_{\bf k}^{(+)}\rangle$. Therefore, we rewrite
the first term by using the definition of the two-body $T$-matrix
$V_{\rm bg}|\psi_{\bf k}^{(+)}\rangle \equiv T_{\rm bg}(2
\epsilon_{\bf k})|{\bf k}\rangle$, where $|{\bf k}\rangle$ denotes
the incoming plane-wave state associated with the scattering state
$|\psi_{\bf k}^{(+)}\rangle$. We thus obtain
\begin{align}
\label{inter1}
&\langle\psi_{\rm m}|V_{\rm am}|\psi_{\bf k}^{(+)}\rangle\nonumber\\
&=\int \frac{d^3{\bf k}'}{(2\pi)^3} \langle\psi_{\rm m}|V_{\rm
am}V_{\rm bg}^{-1}|{\bf k}'\rangle\langle{\bf k}'|T_{\rm
bg}(2\epsilon_{\bf k})|{\bf k}\rangle.
\end{align}
For long wavelengths, the scattering properties of any interatomic
potential become hard-core like, as the incoming particles have
too low energies to probe the structure of the interaction
potential. The background scattering is then described by a
single parameter, the background scattering length $a_{\rm bg}$.
Moreover, the conservation of probability then leads to the simply
result $\langle{\bf k}'|T_{\rm bg}(2\epsilon_{\bf k})|{\bf
k}\rangle \simeq 4\pi a_{\rm bg} \hbar^2/m(1+ika_{\rm bg})$. As a
result, we can introduce the coupling parameter $g$, that
describes the coupling between the two channels by means of
\begin{align}
\label{inter2} \langle\psi_{\rm m}|V_{\rm am}|\psi_{\bf
k}^{(+)}\rangle \equiv \frac{g}{1+ika_{\rm bg}}.
\end{align}
Hence the continuum contribution in the right-hand side of
Eq.~(\ref{2bodyder}) is
\begin{align}
&g^2\int \frac{d^3{\bf k}}{(2\pi)^3}\frac{1}{1+(ka_{\rm bg})^2
}\frac{1}{E-2\epsilon_{\bf k}}.
\end{align}
The zero-energy limit of this expression leads in
Eq.~(\ref{2bodyder}) to a (finite) renormalization of the bare
detuning $\delta_B$ to the experimentally observable detuning
$\delta = \Delta\mu(B-B_0)$ that is exactly zero on resonance.
Here $\Delta\mu$ denotes the magnetic-moment difference between
the open and closed channels of the Feshbach resonance. For $^6$Li
we have $\Delta\mu \simeq 2\mu_B$ in terms of the Bohr magneton
$\mu_B$. Subtracting this limiting value we have that
\begin{align}
\label{sigma2body}
\hbar\Sigma_{\rm m}(E)&=g^2\int \frac{d^3{\bf k}}{(2\pi)^3}\frac{1}{1+(ka_{\rm bg})^2}
\left(\frac{1}{E-2\epsilon_{\bf k}}+\frac{1}{2\epsilon_{\bf k}}\right) \nonumber\\
&=\eta\frac{\sqrt{-E}}{1+\frac{|a_{\rm bg}|\sqrt{m}}{\hbar}\sqrt{-E}},
\end{align}
with $\eta=g^2m^{3/2}/4\pi\hbar^3$ \cite{duine2003}. We have
written $\hbar\Sigma_{\rm m}(E)$ to indicate that this term gives
the self-energy of the molecules, which follows from the fact that
the energy of the exact molecular bound state is found from
\begin{equation}
\label{be}
E=\delta+\hbar\Sigma_{\rm m}(E)~.
\end{equation}

\begin{figure}[htb]
\begin{center}
\includegraphics[width=8cm]{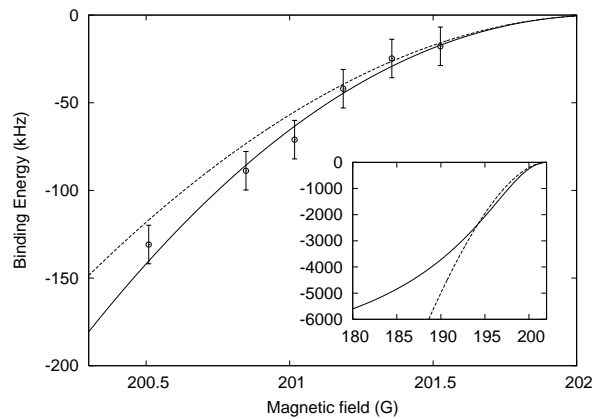}
\end{center}
\caption[a]{Binding energies for the dressed molecules as a
function of magnetic field. The solid line is given by the
solution of Eqs.~(\ref{be}) and (\ref{self}), where we have used
$E_{\rm b} = -8.9 $ MHz for the bound-state energy in the
background potential \cite{nygaard}. The dashed line is the
solution in the absence of background interactions. The points are
from the experiment of Moritz {\it et al.} \cite{esslinger}. The
inset shows how the binding energy behaves as the energy of
the bound state in the background potential is approached and an
avoided crossing occurs.}\label{fig1}
\end{figure}

Now we wish to include also the bound state in the case of a large and
positive background scattering length. For the moment we do not assume
that the bound-state energy lies at $-\hbar^2/m a_{\rm bg}^2$, but
we initially use $E_{\rm b}=-\hbar^2/m a_{\rm b}^2$, where $a_{\rm
b}$ may be different from $a_{\rm bg}$ due to effective-range
effects. To determine the contribution of the bound state, we
compare the normalizations of the wave functions of the bound and
scattering states. We have in the pseudopotential approximation
that
\begin{align}
\label{full2body}
\psi_{{\bf 0}}^{(+)}({\bf r})&=\left(1-\frac{a_{\rm bg}}{r}\right)~, \\
\psi_{\rm b}({\bf r})&=\frac{1}{\sqrt{2\pi a_{\rm b}}}\frac{e^{-r/a_{\rm
b}}}{r}~,
\end{align}
which implies that for small separations between the atoms
\begin{align}
\label{conn} \psi_{\rm b}({\bf r})=-\frac{1}{\sqrt{2\pi a_{\rm
bg}^2a_{\rm b}}}\psi_{{\bf 0}}^{(+)}({\bf r}).
\end{align}
We can now use the results of Eqs.~(\ref{conn}) and (\ref{inter2})
to calculate the self-energy contribution of the bound state.
Subtracting again the zero-energy limit to renormalize the bare
detuning $\delta_B$ we find now
\begin{align}
\label{self}
\hbar\Sigma_{\rm
m}(E)=\eta\left[\frac{\sqrt{-E}}{1+\frac{|a_{\rm
bg}|\sqrt{m}}{\hbar}\sqrt{-E}} -\frac{2\hbar^2/m a_{\rm
bg}^2}{\sqrt{-E_{\rm b}}}\frac{1}{1-E_{\rm b}/E}\right].
\end{align}
Using this result we can show that if $E_{\rm b}=-\hbar^2/m a_{\rm bg}^2$ we
have
\begin{align}
\label{selfminus}
\hbar\Sigma_{\rm m}(E)&=\eta\frac{\sqrt{-E}}{1-\frac{a_{\rm
bg}\sqrt{m}}{\hbar}\sqrt{-E}}.
\end{align}
Note that compared to Eq.~(\ref{sigma2body}) the bound-state
contribution has in the limit of a very large positive background
scattering length resulted in the simple replacement $|a_{\rm bg}|
\rightarrow -a_{\rm bg}$, which shows that the precise value of
the energy of the bound state is all important in determining
whether the combined effect of the scattering continuum and the
bound state shifts the dressed molecular energy up or down. Using
Eqs.~(\ref{be}) and (\ref{self}) we can find the binding energy of
the dressed molecule in the 2-body limit. We have applied this
result to measurements of the binding energy of $^{40}$K molecules
\cite{esslinger} and find excellent agreement with experiment as
shown in Fig.~\ref{fig1}. It is interesting that in this case the
binding energy $-E_{\rm b}$ is substantially larger than
$\hbar^2/m a_{\rm bg}^2$ and it is therefore a better
approximation to use Eq.~(\ref{sigma2body}) for the molecular
self-energy than Eq.~(\ref{selfminus}). The use of the latter
self-energy would shift the molecular binding energies upward
compared to the result with no background interactions, whereas
agreement with the experimental results requires a downward shift.

Now we can go back to the quantity $Z$ that describes the
closed-channel component of the dressed molecule as in
Eq.~(\ref{defz}). While calculating the self-energy of the
molecules, we have used a normalization such that $\langle
\psi_{\rm m}|\psi_{\rm m}\rangle=1$. This implies that the
normalization of the atomic part of the dressed molecular wave
function equals $(1-Z)/Z$, so we have the relation
\begin{align}
\frac{1-Z}{Z}&=\langle \psi_{\rm a}|\psi_{\rm a}\rangle \nonumber\\
&=\left.\langle\psi_{\rm m}|V_{\rm am}\frac{1}{(E-H_{\rm a})^2}
                      V_{\rm am}|\psi_{\rm m}\rangle\right|_{E=\epsilon_{\rm
m}(\delta)}\nonumber\\
&=\left.-\frac{\partial \hbar\Sigma_{\rm m}(E)}{\partial E}\right|_{E=\epsilon_{\rm
m}(\delta)},
\end{align}
where $\epsilon_{\rm m}(\delta)$ is the energy of the exact dressed
molecular state and the solution of the equation $\epsilon_{\rm
m}=\delta+\hbar\Sigma_{\rm m}(\epsilon_{\rm m})$. From this relation it follows that
\begin{align}
\label{Z2body} Z=\left[1-\left.\frac{\partial \hbar\Sigma_{\rm
m}(E)}{\partial E}\right]^{-1}\right|_{E=\epsilon_{\rm
m}(\delta)}.
\end{align}

\begin{figure}[htb]
\begin{center}
\includegraphics[width=8cm]{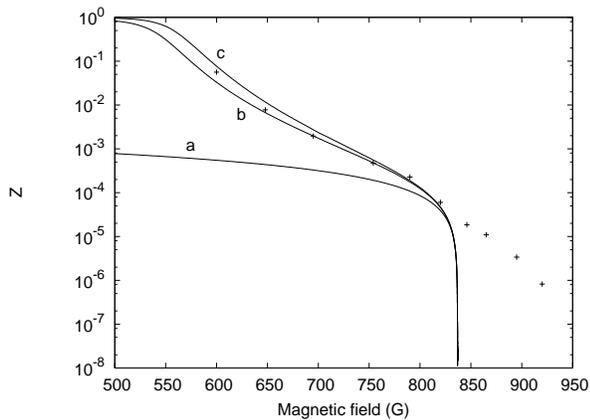}
\end{center}
\caption[a]{
The two-body $Z$ as a function of magnetic field for $^6$Li, (a)
without background interactions, (b) with a constant background
scattering length $a_{\rm bg} = -1360 a_0$ equal to the background
scattering length at resonance, and (c) with a magnetic
field-dependent background scattering length as determined in
Ref.~\cite{falco}. The points are from the experiment of Partridge
{\it et al.} \cite{Rice}.} \label{2bodyfigure}
\end{figure}

The probability
$Z$ of the dressed molecule to be in the closed-channel state was
measured in the experiments of Jochim {\it et al.} \cite{grimm}
and Partridge {\it et al.} \cite{Rice} for $^6$Li. Using the
two-body results derived so far, we can get a very good agreement
on the BEC-side of the resonance, as is shown in
Fig.~\ref{2bodyfigure}. At the two-body level, no stable molecules
exist for positive detuning. Moreover, in that case $Z$ becomes
zero exactly at the resonance. Both features disappear at the
many-body level as we discuss in detail in the next section.

\section{Bogoliubov theory of molecules}
In this section we give a detailed derivation of the Bogoliubov or
RPA theory of the BEC-BCS crossover. We discuss the problems that
arise when trying to numerically calculate the fluctuation
corrections. We also consider the long-wavelength limit of the
theory.
\subsection{Equation of state}
We start the development of the Bogoliubov theory by introducing
the annihilation operator $\psi_{\rm m}({\bf x})$ for the bare
molecules, and the annihilation operators $\psi_{\uparrow}({\bf
x})$ and $\psi_{\downarrow}({\bf x})$ for the atoms in the two
different hyperfine states $|\uparrow\rangle$ and
$|\downarrow\rangle$, respectively. The Hamiltonian can then be
expressed in second quantization as
\cite{drummond,timmermans1999,duine2003}
\begin{align}
\label{hamiltonian} H&=\int d{\bf x}~ \psi^{\dagger}_{\rm m}({\bf
x})\left( -\frac{\hbar^2\mbox{\boldmath $\nabla$}^2}{4m}+
\delta_B-2\mu \right) \psi_{\rm m}({\bf x})
\\ &+ \sum_{\sigma=\uparrow,\downarrow} \int d{\bf x}~
\psi^{\dagger}_{\rm \sigma}({\bf x}) \left(
-\frac{\hbar^2\mbox{\boldmath $\nabla$}^2}{2m}-\mu \right)
\psi_{\rm \sigma}({\bf x}) \nonumber
\\ &+g \int d{\bf x} \left( \psi^{\dagger}_{\rm m}({\bf x})\psi_{\rm
\uparrow}({\bf x})\psi_{\rm \downarrow}({\bf
x})+\psi^{\dagger}_{\rm \downarrow}({\bf x})\psi^{\dagger}_{\rm
\uparrow}({\bf x})\psi_{\rm m}({\bf x}) \right) \nonumber
\\ &+ \frac{4\pi a_{\rm bg} \hbar^2}{m} \int d{\bf x}~ \psi^{\dagger}_{\rm
\downarrow}({\bf x})\psi^{\dagger}_{\rm \uparrow}({\bf
x})\psi_{\rm \uparrow}({\bf x})\psi_{\rm \downarrow}({\bf x})~.
\end{align}
We use functional methods to calculate the physical quantities of interest, as it is simpler than the operator formalism when going beyond mean-field theory. In the end, both the two-body physics, that we treated in the previous section, as well as the many-body BEC-BCS crossover physics are incorporated in our theory. This is achieved by first writing down the grand-canonical partition
function of the gas as the functional integral
\begin{align}
Z_{\rm gr}&=\int d[\psi^*_{\rm m}]d[\psi_{\rm m}]d[\psi^*_{\sigma}]d[\psi_{\sigma}]
 e^{-S[\psi^*_{\rm m},\psi_{\rm m},\psi_{\sigma}^*,\psi_{\sigma}]/\hbar},
\end{align}
where the Euclidean or imaginary-time action is
\begin{align}
S&[\psi^*_{\rm m},\psi_{\rm m},\psi_{\sigma}^*,\psi_{\sigma}]
=\int_0^{\hbar\beta}d\tau
\Bigg \{ H[\psi^*_{\rm m},\psi_{\rm m},\psi_{\sigma}^*,\psi_{\sigma}]
\nonumber \\ &+
\sum_{\sigma=\uparrow,\downarrow} \int d{\bf x}~
\psi^*_{\sigma} ({\bf x},\tau) \hbar\frac{\partial}{\partial\tau}\psi_{\sigma}({\bf x},\tau)
\nonumber\\ 
&+ \int d{\bf x}~
\psi^*_{\rm m}({\bf x},\tau) \hbar\frac{\partial}{\partial\tau}\psi_{\rm m}({\bf x},\tau)
\Bigg \}.
\end{align}
We then rewrite the open-channel part of this action using Green's function methods, where the Green's function $G({\bf x},\tau;{\bf x}',\tau')$ can be seen as a `matrix' with space and time `indices'. We obtain
\begin{align}
S&=\int \! d\tau\,d{\bf x}\:
\psi^*_{\rm m}({\bf x},\tau)\!
\left(\hbar\frac{\partial}{\partial\tau}-\frac{\hbar^2\nabla^2}{4m}
  +\delta_B-2\mu\right)\!\psi_{\rm m}({\bf x},\tau)\nonumber\\
&-\int \! d\tau\,d{\bf x} \, d\tau'\,d{\bf x}'\:
\left[\psi^*_{\downarrow}({\bf x},\tau),\psi_{\uparrow}({\bf x},\tau)\right]
\hbar G^{-1}
\left[\begin{matrix}
\psi_{\downarrow}({\bf x}',\tau')\\
\psi^*_{\uparrow}({\bf x}',\tau')
\end{matrix}\right].
\label{action}
\end{align}
The coupling between the atoms and the molecules is contained in the
atomic $2 \times 2$ Nambu-space matrix propagator $G({\bf
x},\tau;{\bf x}',\tau')$, i.e., $G=G[\psi^*_{\rm m},\psi_{\rm m}]$
depends on the molecular field $\psi_{\rm m}({\bf x},\tau)$. This is a convenient rewrite, as it will allow us to integrate out the atoms using Gaussian integration. It also makes a clear connection with the operator formalism, as the propagator is related to the expectation value of the imaginary-time Heisenberg operators in the following manner
\begin{align}
G&({\bf x},\tau;{\bf x}',\tau')
\nonumber\\
& = -\left \langle T \left (
\left[\begin{matrix}
\psi_{\downarrow}({\bf x},\tau)\\
\psi^{\dagger}_{\uparrow}({\bf x},\tau)
\end{matrix}\right]
\left[\psi^{\dagger}_{\downarrow}({\bf x}',\tau'),\psi_{\uparrow}({\bf x}',\tau')\right]
\right)
\right \rangle,
\label{connection}
\end{align}
where $T$ is the time-ordering operator. The 2x2 matrix-structure allows us for including the coupling to the molecules and in particular immediately incorporates the anomalous expectation values of the atomic gas that arise when the molecules Bose-Einstein condense.

\begin{figure}[h]
\begin{center}
\includegraphics[width=3cm]{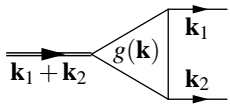}
\end{center}
\caption[a]{
Momentum-dependent atom-molecule coupling. The incoming molecular
field is represented by the double lines and the single lines
represent the outgoing atoms. The coupling $g$ depends on the
relative momentum ${\bf k}=({\bf k}_2-{\bf k}_1)/2$ of the
outgoing atoms in the case of a non-negligible background
interaction.} \label{g-diagram}
\end{figure}

It appears that we have neglected
here the background interaction between the atoms. However, we
reintroduce the effects of the background scattering on the
coupling $g$ once we go back to momentum space, which, in
agreement with Eq.~(\ref{inter2}), results in the substitution $g \rightarrow
g/(1+ika_{\rm bg})$ where $k$ is the magnitude of the relative
momentum between the two atoms. Diagrammatically this procedure is
shown in Fig.~\ref{g-diagram}.

\begin{figure}[htb]
\begin{center}
\includegraphics[width=7cm]{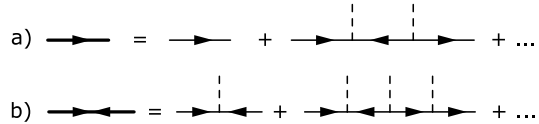}
\end{center}
\caption[a]{
Diagrammatic expansion of the normal and anomalous BCS propagators. The thick lines represent the full BCS propagator, the thin lines represent the bare propagator. Interactions with the molecular Bose-Einstein condensate are denoted by the dashed lines. a) The normal BCS propagator. b) The anomalous propagator.}
\label{propa}
\end{figure}

Our aim is to exactly integrate out the atoms and expand the
result up to second order in the fluctuations of the molecular
field around its expectation value $\langle \psi_{\rm m}({\bf
x},\tau) \rangle$, which we assume to be real from now on, and
thus arrive at a Bogoliubov theory for the bare molecules. To make
the connection with BCS-theory explicit, we write $\langle
\psi_{\rm m}({\bf x},\tau) \rangle = \Delta/g$, perform the
substitution $\psi_{\rm m}({\bf x},\tau) \rightarrow \Delta/g +
\psi_{\rm m}({\bf x},\tau)$, and use the Dyson equation
\begin{align}
G^{-1}=G_{\rm a}^{-1}-\Sigma
\end{align}
with
\begin{align}
G_{\rm a}^{-1} =&
-\frac{1}{\hbar}
\left(
\begin{matrix}
 \hbar\frac{\partial}{\partial\tau}-\frac{\hbar^2\nabla^2}{2m}-\mu & \Delta\\
\Delta &
\hbar\frac{\partial}{\partial\tau}+\frac{\hbar^2\nabla^2}{2m}+\mu
\end{matrix}
\right)\nonumber\\
&\times\delta({\bf x}-{\bf x}')\delta(\tau - \tau'),
\nonumber\\
\Sigma=&\frac{1}{\hbar}
\left( \begin{array}{cc}
0 & g \psi_{\rm m}({\bf x},\tau) \\
g \psi_{\rm m}^*({\bf x},\tau) & 0 \\
\end{array} \right)\delta({\bf x}-{\bf x}')\delta(\tau - \tau')~.
\label{BCSpropa}
\end{align}
The reason for this procedure is that $G^{-1}_{\rm a}$ is now exactly the usual BCS propagator, which we know how to deal with exactly from BCS theory. The `gap' $\Delta$ is in our case proportional to the expectation value of the molecular field, i.e., it represents the molecular BEC. Two atoms can be annihilated through the formation of a bare molecule in the condensate. Vice-versa, two atoms can also be created from the molecular condensate. Diagrammatically, the propagator can be expanded as is shown in Fig.~\ref{propa}.

\begin{figure}[htb]
\begin{center}
\includegraphics[width=5cm]{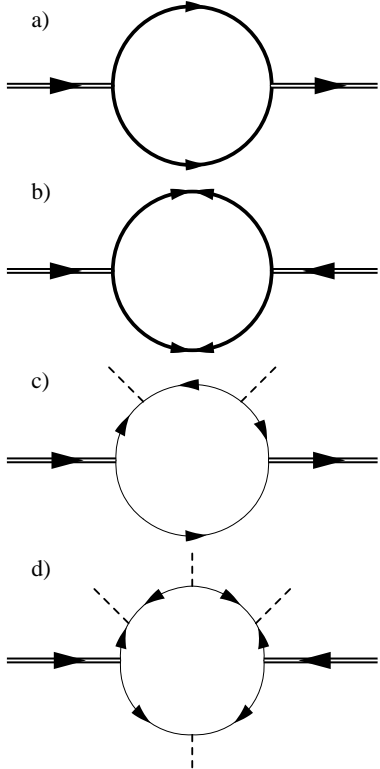}
\end{center}
\caption[a]{
Diagrammatic representation of the molecular self-energy diagrams.
The bare fermionic propagators are dressed through interactions
with the molecular condensate. Single lines denote fermions,
double lines denote molecules. For the fermions, the thin lines
represent the bare propagator, thick lines represent the BCS
propagator, as in Fig.~\ref{propa}. Interactions with the condensate are denoted by the
vertices containing a dashed line. The effect of the background
scattering on the vertices is not drawn explicitly. a) The normal
self-energy $\hbar\Sigma_{11}$. b) The anomalous self-energy
$\hbar\Sigma_{12}$. c) An example of a contribution to the normal
self-energy $\hbar\Sigma_{11}$. d) An example of a contribution to the anomalous
self-energy $\hbar\Sigma_{12}$.} \label{diagrams}
\end{figure}

The self-energy term $\Sigma$ accounts for the molecular fluctuations, and as such depends on the molecular field, $\Sigma=\Sigma[\psi^*_{\rm
m},\psi_{\rm m}]$. The action has become quadratic in the atomic fields, and so we can now formally integrate out the atoms by doing the Gaussian integral. We make use of the fact that $G$ can be seen as a matrix with space-time `indices' and obtain
\begin{align}
Z_{\rm gr}=&\int d[\psi^*_{\rm m}]d[\psi_{\rm m}]
\exp\Big\{-S_{\rm m}[\psi^*_{\rm m},\psi_{\rm m}]/\hbar
\nonumber \\
&+{\rm Tr}\left[\ln(-G^{-1}[\psi^*_{\rm
m},\psi_{\rm m}])\right]\Big\},
\end{align}
where the trace is over Nambu space, and over the space and
(imaginary) time coordinates. We next expand this exact result up
to second-order in the molecular fluctuations by means of
\begin{align}
&{\rm Tr}[\ln(-G^{-1})]
={\rm Tr}[\ln(-G_{\rm a}^{-1}(1-G_{\rm a}\Sigma))]\nonumber\\
&\phantom{blad}={\rm Tr}[\ln(-G_{\rm a}^{-1})]-{\rm Tr}[G_{\rm
a}\Sigma +\frac{1}{2}G_{\rm a}\Sigma G_{\rm a}\Sigma+\dots].
\end{align}
Since the self-energy $\hbar\Sigma$ is linear in $\psi_{\rm m}$
this is indeed effectively an expansion in the molecular
fluctuations. The first-order term, ${\rm Tr}[G_{\rm a} \Sigma]$,
is required to be equal to $(2\mu-\delta_B)(\psi_{\rm
m}+\psi^*_{\rm m})/\hbar$ because we want to expand around the
minimum of the effective action. Using the expressions for
$\Sigma$ and $G_{\rm a}$, and going over to momentum space, the
second-order term can be written in the zero-temperature limit as
\begin{widetext}
\begin{align}
\label{self-energies}
{\rm Tr}\left[G_{\rm a} \Sigma G_{\rm a} \Sigma\right]=&
\sum_{{\bf k},n}
\left[
\psi_{\rm m}^*({\bf k},i\omega_n),\psi_{\rm m}(-{\bf k},-i\omega_n)
\right]
\left[\begin{array}{cc}
\Sigma_{11}({\bf k},i\omega_n) & \Sigma_{12}({\bf k},i\omega_n) \\
\Sigma_{21}({\bf k},i\omega_n) & \Sigma_{22}({\bf k},i\omega_n) \\
\end{array}\right]
\left[\begin{array}{c}
\psi_{\rm m}({\bf k},i\omega_n) \\
\psi^*_{\rm m}(-{\bf k},-i\omega_n) \\
\end{array}\right].
\end{align}
For example, $\Sigma_{11}$ contains two normal fermionic propagators, which can be taken from Eq.~(\ref{BCSpropa}). In momentum space, this ultimately leads to
\begin{align}
\hbar\Sigma_{11}({\bf k},i \omega_n)=-\frac{1}{\beta} \int \frac{d{\bf k'}}{(2\pi)^3}~ \sum_{n'}|g({\bf k}')|^2
\frac{i \hbar\omega_{n'_+}\!+\epsilon({\bf k}'_+)-\mu}
{[\hbar\omega_{n'_+}]^2+[\hbar\omega_a({\bf k}'_+)]^2}
\frac{i \hbar\omega_{n'_-}\! +\epsilon({\bf k}'_-)-\mu}
{[\hbar\omega_{n'_-}]^2+[\hbar\omega_a({\bf k}'_-)]^2}
\end{align}
Here we have also introduced the BCS dispersion $\hbar\omega_{\rm
a}({\bf k})=[(\epsilon({\bf k})- \mu)^2 + \Delta^2/(1+a_{\rm bg}^2
{\bf k}^2)]^{1/2}$, the atomic dispersion $\epsilon({\bf
k})=\hbar^2{\bf k}^2/2m$, and the notation ${\bf k}'_{\pm}={\bf k}/2
\pm {\bf k}'$ and $n'_{\pm}=n/2
\pm n'$. The coupling $g$ is dressed by the background scattering
through $|g({\bf k})|^2=g^2/(1+{\bf k}^2a_{\rm bg}^2)$ as
explained previously. The two propagators can be reduced using partial fraction decomposition, resulting in the expression
\begin{align}
\hbar\Sigma_{11}({\bf k},i \omega_n)=\frac{1}{\beta} \int \frac{d{\bf k'}}{(2\pi)^3}~ \sum_{n'}|g({\bf k}')|^2
\Bigg\{
\frac{u_{\rm a}^2({\bf k}'_+) u_{\rm a}^2({\bf k}'_-)}
     {\hbar\omega_{\rm a}({\bf k}'_+)+\hbar\omega_{\rm a}({\bf k}'_-)-i\hbar\omega_n}
&\left[\frac{1}{i \hbar\omega_{n'_+}\!-\hbar\omega_{\rm a}({\bf k}'_+)}
+\frac{1}{i \hbar\omega_{n'_-}\!-\hbar\omega_{\rm a}({\bf k}'_-)}\right]
\nonumber\\
+\frac{u_{\rm a}^2({\bf k}'_+) v_{\rm a}^2({\bf k}'_-)}
     {\hbar\omega_{\rm a}({\bf k}'_+)-\hbar\omega_{\rm a}({\bf k}'_-)-i\hbar\omega_n}
&\left[\frac{1}{i \hbar\omega_{n'_+}\!-\hbar\omega_{\rm a}({\bf k}'_+)}
+\frac{1}{i \hbar\omega_{n'_-}\!+\hbar\omega_{\rm a}({\bf k}'_-)}\right]
\nonumber\\
+\frac{v_{\rm a}^2({\bf k}'_+)u_{\rm a}^2({\bf k}'_-)}
     {-\hbar\omega_{\rm a}({\bf k}'_+)+\hbar\omega_{\rm a}({\bf k}'_-)-i\hbar\omega_n}
&\left[\frac{1}{i \hbar\omega_{n'_+}\!+\hbar\omega_{\rm a}({\bf k}'_+)}
+\frac{1}{i \hbar\omega_{n'_-}\!-\hbar\omega_{\rm a}({\bf k}'_-)}\right]
\nonumber\\
+\frac{v_{\rm a}^2({\bf k}'_+) v_{\rm a}^2({\bf k}'_-)}
     {-\hbar\omega_{\rm a}({\bf k}'_+)-\hbar\omega_{\rm a}({\bf k}'_-)-i\hbar\omega_n}
&\left[\frac{1}{i \hbar\omega_{n'_+}\!+\hbar\omega_{\rm a}({\bf k}'_+)}
+\frac{1}{i \hbar\omega_{n'_-}\!+\hbar\omega_{\rm a}({\bf k}'_-)}\right]
\Bigg\},
\label{partsigma11}
\end{align}
\end{widetext}
where $u_{\rm a}({\bf k})$ and $v_{\rm a}({\bf k})$ are the usual BCS coherence factors, obeying
\begin{align}
u_{\rm a}^2({\bf k})=\frac{\hbar\omega_{\rm a}({\bf
k})+\epsilon({\bf k})-\mu} {2\hbar\omega_{\rm
a}({\bf k})}, \\ 
v_{\rm a}^2({\bf k})=\frac{\hbar\omega_{\rm a}({\bf
k})-\epsilon({\bf k})+\mu} {2\hbar\omega_{\rm a}({\bf k})}.
\end{align}
From the terms in the right-hand side of Eq.~(\ref{partsigma11}), we can read off the different processes that lead to a dressing of the bare molecules. The ladder diagram which this expression corresponds to is shown in Fig.~\ref{diagrams}a. It describes two fermions, propagating with momentum ${\bf k}_+$ and ${\bf k}_-$. In the BCS-picture, there are indeed four possible processes that renormalize the molecules, due to the possibility for the molecules to break up into two quasi-particles (proportional to $u^2_{\rm a}u^2_{\rm a}$), a quasi-hole and a quasi-particle (proportional to $u^2_{\rm a}v^2_{\rm a}$), or two quasi-holes (proportional to $v^2_{\rm a}v^2_{\rm a}$).

Next, we take the zero-temperature limit by replacing the sum over internal Matsubara frequencies $n'$ by an integral. As a result, only two processes survive: the creation of a quasi-particle pair, and the creation of a quasi-hole pair. Finally, we perform the same renormalization of the detuning as explained in Sec.~II. In this manner we are ensured that $\delta=0$ corresponds also in the many-body theory exactly to the location of the Feshbach resonance. To conclude, the self-energy contributions are now
\begin{widetext}
\begin{eqnarray}
\label{self-energy-elements}
\hbar\Sigma_{11}({\bf k},i\omega_n) &=&
     \int \frac{d{\bf k'}}{(2\pi)^3}~|g({\bf k'})|^2
     \Bigg\{\frac{u_{\rm a}^2({\bf k}'_+) u_{\rm a}^2({\bf k}'_-)}
       {i\hbar\omega_n-\hbar\omega_{\rm a}({\bf k}'_+)-\hbar\omega_{\rm a}({\bf k}'_-)}-
     \frac{v_{\rm a}^2({\bf k}'_+) v_{\rm a}^2({\bf k}'_-)}
       {i\hbar\omega_n+\hbar\omega_{\rm a}({\bf k'}_+)+\hbar\omega_{\rm a}({\bf k}'_-)}+
     \frac{1}{2 \epsilon({\bf k}')}\Bigg\}~, \nonumber\\
\hbar\Sigma_{12}({\bf k},i\omega_n) &=& 2
     \int \frac{d{\bf k}'}{(2\pi)^3}~|g({\bf k}')|^2 \Bigg\{
       u_{\rm a}({\bf k}'_+)v_{\rm a}({\bf k}'_+)u_{\rm a}({\bf k}'_-)v_{\rm a}({\bf k}'_-)
     \frac{\hbar\omega_{\rm a}({\bf k}'_+)+\hbar\omega_{\rm a}({\bf k}'_-)}
       {(\hbar\omega_{\rm a}({\bf k}'_+)+\hbar\omega_{\rm a}({\bf k'}_-))^2
           +(\hbar\omega_n)^2}\Bigg\}~, \nonumber\\
\hbar\Sigma_{22}({\bf k},i\omega_n) &=& \hbar\Sigma_{11}({\bf k},-i\omega_n), \nonumber\\
\hbar\Sigma_{21}({\bf k},i\omega_n) &=& \hbar\Sigma_{12}({\bf k},-i\omega_n).
\end{eqnarray}
\end{widetext}
 
In our calculations, we also take into account that the
background scattering length $a_{\rm bg}(B)$ and the coupling
$g(B)$ depend on the magnetic field, which turns out to be very
important for the extremely broad Feshbach resonance used in all
experiments with $^6$Li \cite{falco}.

Diagrammatically, Fig.~\ref{diagrams}
illustrates the way the BCS propagators are included into the
theory. The ladder diagram in a) is the normal self-energy $\hbar\Sigma_{11}$. The BCS-propagators can be expanded as is shown in Fig.~\ref{propa}, which gives rise to contributions such as the one in c). The diagram in b) is the anomalous self-energy $\hbar\Sigma_{12}$. A contribution using bare propagators is shown in d). This shows explicitly that there is only an anomalous contribution for nonzero $\Delta$.

In terms of the above self-energies, the bare molecular
condensate density $\Delta/g$ is determined by the exact
Hugenholtz-Pines relation $2\mu = \delta + \hbar\Sigma_{11}({\bf
0},0)- \hbar\Sigma_{12}({\bf 0},0)$ \cite{hugpines}, which in our case turns
out to be equal to the modified BCS gap equation
\begin{eqnarray}
\label{eqgap}
\delta - 2\mu = \int \frac{d{\bf k}}{(2\pi)^3}~ |g({\bf k})|^2
\left( \frac{1}{2\hbar\omega_{\rm a}({\bf k})} - \frac{1}{2\epsilon({\bf
k})} \right).
\end{eqnarray}
As expected, the same gap equation can also be obtained from the
requirement that the linear terms $(\delta_B-2\mu)(\psi_{\rm
m}+\psi^*_{\rm m})+{\rm Tr}[G_{\rm a} \hbar\Sigma]$ in the
effective molecular action vanish.

Integrating out the atoms has led to an effective Bogoliubov
action for the bare molecules. The interaction effects between the
molecules and atoms are contained in the normal and anomalous
self-energies $\hbar\Sigma_{ij}$. Because we have expanded in the
molecular fluctuations only, the BCS propagators for the atomic
fields are included everywhere, and the BCS state is fully
incorporated in the physics. Beyond that, we have constructed an effective action that
also describes the fluctuations or noncondensed molecules, which
serves to draw not only conclusions about the condensate depletion
but also on the nature of the dressed molecules in the condensate
as we will see in a moment.

The inverse propagator of the molecules takes the form
\begin{align}
-\hbar G_{\rm m}^{-1}=
\left[\begin{array}{cc}
-\hbar G_{11}^{-1}+\hbar \Sigma_{11} & \hbar \Sigma_{12}\\
\hbar \Sigma_{21} & -\hbar G_{22}^{-1}+\hbar \Sigma_{22} \\
\end{array}\right],
\end{align}
where
\begin{align}
-\hbar G_{11}^{-1}({\bf k},i\omega)&=-i \hbar \omega+\epsilon({\bf k})/2+\delta-2\mu,
                                                                             \nonumber\\
-\hbar G_{22}^{-1}({\bf k},i\omega)&=\phantom{-}i \hbar \omega
 +\epsilon({\bf k})/2+\delta-2\mu
\end{align}
are taken from Eq.~(\ref{action}). The molecular propagator can be written as the expectation value of time-ordered molecular creation and annihilation operators, similarly as the atomic propagator in Eq.~(\ref{connection}).
Using these expressions we can obtain the grand-canonical partition function
$Z_{\rm gr}$ and the thermodynamic potential $\Omega=-\ln(Z_{\rm
gr})/\beta$ in the Bogoliubov approximation by performing the
Gaussian integral over the molecular fluctuations. Ultimately we
obtain in this manner
\begin{align}
Z_{\rm gr}=&\exp\bigg\{
{\rm Tr}\left[\ln(-G_{\rm a}^{-1})\right]-
\frac{1}{2}{\rm Tr}\left[\ln(-G_{\rm m}^{-1})\right]
\nonumber\\
&\phantom{\exp\bigg\{}-\beta(\delta-2\mu)V\frac{|\Delta|^2}{g^2}\bigg\},
\end{align}
where $V$ is the volume of the gas. As the molecular and atomic
contributions have now mixed, it is appropriate to determine the
equation of state by taking the derivative of the thermodynamic
potential with respect to the chemical potential. Hence, the total
atomic density becomes
\begin{align}
\label{eqst}
n=&-\frac{1}{V}\frac{\partial\Omega}{\partial\mu}\nonumber\\
=& \frac{1}{\hbar\beta V}{\rm Tr}\left[ G_{\rm
a}\sigma_3\right]+\frac{2\Delta^2}{g^2}-\frac{1}{2 \beta V}{\rm
Tr}\left[ \frac{G_{\rm m}}{\hbar}\frac{\partial\hbar G_{\rm
m}^{-1}}{\partial\mu}\right]
\nonumber \\
=&\frac{2}{V}\sum_{\bf k}|v_{\rm a}({\bf
k})|^2+\frac{2\Delta^2}{g^2} -\frac{1}{\hbar \beta V}{\rm
Tr}\left[G_{\rm m}\right]
\nonumber\\
&+\frac{1}{2 \hbar \beta V}{\rm Tr}\left[ G_{\rm
m}\frac{\partial\hbar \Sigma}{\partial\mu}\right] .
\end{align}
Here the third Pauli matrix in Nambu space is denoted by
$\sigma_3$. Because we are working at zero temperature and in the
thermodynamic limit, we can rewrite the trace as an integral over
momenta and frequencies. In detail this means that $\sum_{{\bf
k},n}/\hbar \beta V = \int d\omega d{\bf k}\,/(2 \pi)^4$.

The four terms in the right-hand side of Eq.~(\ref{eqst}) have the
following physical interpretation. The first term is the usual BCS
expression for the density of atoms in the BCS ground state. The second term
gives the contribution of the bare molecular Bose-Einstein
condensate to the total atomic density, and the third term gives
the contribution of the bare molecules that are not Bose-Einstein
condensed, i.e., it describes the depletion of the bare molecular
Bose-Einstein condensate. The fourth and last term is most
difficult to understand and can be best explained by reformulating
the equation of state in terms of dressed molecules instead of
bare molecules. If the gas contains a Bose-Einstein condensate of
dressed molecules, there is both a bare molecular contribution and
an atomic contribution to the total atomic density of the
Bose-Einstein condensate. These two contributions, together with
the contribution of the unpaired atoms, are contained in the first
two terms in the right-hand side of Eq.~(\ref{eqst}). These two
terms correspond to the mean-field theory of the BEC-BCS
crossover. The third and the fourth terms in the right-hand side of
Eq.~(\ref{eqst}) represent the effects of fluctuations. Together
they give essentially the bare molecular contribution and the
atomic contribution to the total atomic density of the dressed
molecules that are not Bose-Einstein condensed. In the Bogoliubov
theory, therefore, the gas consists of unpaired atoms,
Bose-Einstein condensed dressed molecules, and dressed molecules
that are not Bose-Einstein condensed.

\subsection{Numerical methods}
Though the analytical expressions for our theory are quite compact
and their behavior can be studied in all limits, solving the gap
equation in Eq.~(\ref{eqgap}) and the equation of state in
Eq.~(\ref{eqst}) simultaneously is quite involved. It is important
to examine these equations carefully, in order to avoid numerical
difficulties. It should be noted that the integrand of the
self-energy not only depends on the magnitudes of the internal and
external momenta, but also on the angle between them, so the
integration over internal momenta is in fact a two-dimensional
one. In combination with the integration over the external momenta
and Matsubara frequencies and the iteration procedure required to
find a self-consistent solution at all magnetic fields, this leads
to the practical necessity of choosing a fast method of
calculation where possible. Convergence issues turn also out to be
a problem, especially in the case of a broad resonance.

Our general procedure is to solve the gap equation and the
equation of state iteratively, to find the desired value of the
total atomic density. The input for this procedure is the gap
$\Delta$, which is completely fixed during the iterations. Then we
take a chemical potential, and solve the gap equation to find the
detuning. Finally, the equation of state gives the atomic density
that we can compare with the desired experimental density. We
adjust the chemical potential appropriately, and start the next
iteration. We perform this procedure for different $\Delta$ to
obtain a self-consistent solution for various values of the
magnetic field. The reason why we keep $\Delta$ fixed rather than
the detuning, is that this ensures that a physical solution of the
gap equation can always be found.

\begin{figure}[htb]
\begin{center}
\includegraphics[width=6cm]{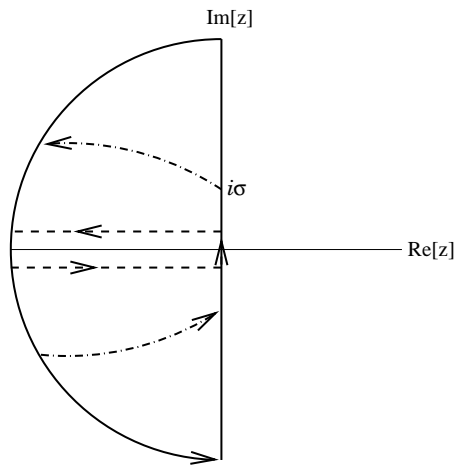}
\end{center}
\caption[a]{
Tracing the Matsubara frequencies using various methods of contour integration.}
\label{contour}
\end{figure}

We have found that the procedure of tracing over the Matsubara
frequencies is very subtle. The first possibility of doing this is
to just replace the discrete sum by an integral, due to the
zero-temperature limit, and to perform a straightforward
integration over the frequencies along the imaginary axis of the
complex frequency plane. This is shown in Fig.~\ref{contour} by
the solid curve. In this figure, the frequencies are shown as a
complex energy $z=i \hbar \omega_n$. Using this method, we
calculate the equal-time propagators $\sum_n G({\bf k},i
\omega_n)=N({\bf k})+1/2$, and therefore a convergence factor is
needed to obtain the correct occupation number $N({\bf k})$. This
is equivalent to closing the contour on the left with a large
semicircle.

For the third and fourth terms in the right-hand side of
Eq.~(\ref{eqst}) this method has a significant disadvantage. In
particular, for the third term and for high external momenta, the
integration over the imaginary axis, and the closing of the
contour add to zero, and this means the calculation would be very
sensitive to the precision with which these contributions can be
calculated.

An alternative is to perform a contour deformation, such that the
path of integration lies infinitesimally below and above the real
axis, as is shown by the dashed lines in Fig.~\ref{contour}. This
is usually a desirable method, as there is no need to correct for
the value of the equal-time propagators. The use of this method,
however, necessitates knowledge of the analytic structure of the
retarded propagator to accurately perform the numerical
integration, which is not feasible to obtain from our complicated
expressions for the self-energy. Therefore, we prefer to use
another method by integrating over some intermediate path, and to
work with complex energies along the integration. Any parabolic
shape would seem very suitable, as it also takes away the need for
closing the contour on the left side. Unfortunately, these curves
have a strong positive and negative contribution over a wide range
of energies, that would again have to cancel.

For the third term in the right-hand side of Eq.~(\ref{eqst}), the
best approach, shown by the dash-dotted line in
Fig.~\ref{contour}, is to go along the imaginary axis for a finite
interval $\sigma$, and then to follow a special path given by
\begin{align}
z(p)=&\epsilon({\bf k}) /2-2 \mu\nonumber\\
&- \left[(p+1)\sqrt{\epsilon({\bf k}) /2-2 \mu -i \sigma }+p \eta\right]^2,
\end{align}
where $p$ parametrizes the path and $\epsilon({\bf k})$ is the
external momentum. This expression is chosen such that the
two-body self-energy would give exactly zero along the path.
Closing the contour still gives a correction of
\begin{align}
\frac{4}{\pi } {\rm Arg}\left[\eta +\sqrt{\epsilon({\bf k})/2-2 \mu -i \sigma }\right],
\end{align}
where ${\rm Arg[z]}$ gives the argument of the complex number $z$.
This term becomes small for small $\sigma$ and large
$\epsilon({\bf k})$. For the fourth term in the right-hand side of
Eq.~(\ref{eqst}), such a procedure is not possible, and we resort
to a parabolic path around the negative real axis. In this case,
the convergence is better, as the integrand behaves as $1/|z|^2$.

\subsection{Long-wavelength limit}

We finally study the behavior of the poles of the bare molecular
propagator. The molecular propagator at zero momentum describes
the closed-channel part of the Bose-Einstein condensate of dressed
molecules and therefore has a pole at zero frequency. The strength of this pole is of great interest to us, as we will see in a moment.

The behavior of the propagator for small ${\bf k}$ and $\omega$
can be studied by expanding around this point. We have
\begin{align}
\hbar\Sigma_{11}({\bf k},z)&=\hbar\Sigma_{11}^{(0,0)}
+\Sigma_{11}^{(0,1)} z+\frac{1}{2}\Sigma_{11}^{(0,2)} z^2 + \nonumber \\
&\phantom{blabla}\Sigma_{11}^{(1,0)}\epsilon({\bf k})+\ldots\nonumber \\
\hbar\Sigma_{12}({\bf k},z)&=\hbar\Sigma_{12}^{(0,0)}+\frac{1}{2}\Sigma_{12}^{(0,2)} z^2
+\nonumber \\
&\phantom{blabla}\Sigma_{12}^{(1,0)}\epsilon({\bf k})+\ldots,
\end{align}
where the following definitions have been used
\begin{equation}
\begin{split}
\label{defcoef}
\hbar\Sigma_{11}^{(0,0)}&=\hbar\Sigma_{11}({\bf 0},0)\\
\hbar\Sigma_{12}^{(0,0)}&=\hbar\Sigma_{12}({\bf 0},0)\\
\hbar\Sigma_{11}^{(0,1)}&=\left.\frac{\partial}{\partial z}\hbar\Sigma_{11}({\bf k},z)\right|_{{\bf k}={\bf 0},z=0}\\
\hbar\Sigma_{11}^{(0,2)}&=\left.\frac{\partial^2}{\partial z^2}\hbar\Sigma_{11}({\bf k},z)\right|_{{\bf k}={\bf 0},z=0}\\
\hbar\Sigma_{11}^{(1,0)}&=\left.\frac{\partial}{\partial \epsilon({\bf k})}\hbar\Sigma_{11}({\bf k},z)\right|_{{\bf k}={\bf 0},z=0}\\
\hbar\Sigma_{12}^{(1,0)}&=\left.\frac{\partial}{\partial \epsilon({\bf k})}\hbar\Sigma_{12}({\bf k},z)\right|_{{\bf k}={\bf 0},z=0}\\
\hbar\Sigma_{12}^{(0,2)}&=\left.\frac{\partial^2}{\partial z^2}\hbar\Sigma_{12}({\bf k},z)\right|_{{\bf k}={\bf 0},z=0}.
\end{split}
\end{equation}
We can find analytic expressions for all of these coefficients
using Eq.~(\ref{self-energy-elements}). 

Now we introduce the spectral function, by Wick-rotating the molecular propagator to real energies, i.e., $G_{{\rm m},11}({\bf k},i\omega) \rightarrow G_{{\rm m},11}({\bf k},\omega^+)$. The imaginary part of this expression gives the spectral function of the bare molecules. This spectral function gives us the wave-function overlap between a bare molecule and the two-atom elementary excitation of our many-body system. For comparison, in the limit of two-body physics, the spectral function can be defined in a similar way. It is given by
\begin{align}
\rho&_{\rm m}({\bf k},\omega)
\nonumber\\
&=-\frac{1}{\pi} {\rm Im} \left[\frac{1}{\hbar \omega^+ -\epsilon({\bf k})/2-\delta-\hbar\Sigma_{\rm m}(\hbar\omega^+-\epsilon({\bf k})/2)}\right],
\end{align} 
with the self-energy given in Eq.~(\ref{sigma2body}). For negative detuning, this spectral function contains a delta peak at negative frequencies, corresponding to the dressed molecular state, and a tail for positive frequencies associated with the two-body scattering states. As the delta peak describes the overlap between the dressed and molecular state, its strength is exactly equal to $Z$. 

In the many-body case, the situation is similar. At zero momentum, the delta peak lies at zero energy, because it is the probability of creating a dressed molecule in the Bose-Einstein condensate. Its strength is equal to the sought-after factor $Z$. We derive analytically that 
\begin{align}
\label{peaks}
&\rho_{\rm m}({\bf k},\omega) \equiv -\frac{1}{\hbar\pi} {\rm Im}\left[G_{11}({\bf k},\omega^+)\right]
\simeq\nonumber\\
&+\left(
\frac{1-\hbar\Sigma_{11}^{(0,1)}}{2\alpha^2}+
\frac{|\hbar\Sigma_{12}^{(0,0)}|^{1/2}}{2 \alpha \sqrt{\epsilon({\bf k})}}
\right)
\delta\left(\hbar\omega-
\hbar c k \right)
\nonumber \\
&+\left(
\frac{1-\hbar\Sigma_{11}^{(0,1)}}{2\alpha^2}-
\frac{|\hbar\Sigma_{12}^{(0,0)}|^{1/2}}{2 \alpha \sqrt{\epsilon({\bf k})}}
\right)
\delta\left(\hbar\omega+
\hbar c k \right),
\end{align}
with $\alpha^2=(1-\hbar\Sigma_{11}^{(0,1)})^2+|\hbar\Sigma_{12}^{(0)}|(\hbar\Sigma_{12}^{(0,2)}-\hbar\Sigma_{11}^{(0,2)})$. The speed of sound is given by
\begin{align}
\label{sound}
c=\left[\frac{1+2\hbar\Sigma_{11}^{(1,0)}-2\hbar\Sigma_{12}^{(1,0)}}{2m \left[\hbar\Sigma_{12}^{(0,2)}-\hbar\Sigma_{11}^{(0,2)}+\left(1-\hbar\Sigma_{11}^{(0,1)}\right)^2/\hbar\Sigma_{12}^{(0)}\right]}\right]^{1/2}.
\end{align}
This shows that our theory has reduced to a renormalized
Bogoliubov-theory for the bare molecules. The delta peak at
positive energy corresponds to a particle-like excitation, and its
strength is equal to the Bogoliubov amplitude $Z|u_{\rm m}({\bf
k})|^2$. The delta peak at negative-energy corresponds to a
hole-like excitation and has strength $Z|v_{\rm m}({\bf k})|^2$.
Both of these are renormalized by an overall factor $Z$, but from
their separate expressions this is not immediately clear. However,
as mentioned before, we expect the combined effect of the two
delta peaks at zero momentum to result in a single delta peak in
the spectral function with a strength $Z$. From the expression in
Eq.~(\ref{peaks}) this is found to be the case, and the factor $Z$
is found to be equal to
\begin{align}
\label{Z}
Z=\frac{1-\Sigma_{11}^{(0,1)}}{(1-\Sigma_{11}^{(0,1)})^2+|\hbar\Sigma_{12}^{(0)}|(\Sigma_{12}^{(0,2)}-\Sigma_{11}^{(0,2)})}.
\end{align}

\section{Broad resonances}
In the case of a broad resonance, such as the one at 834 G for
$^6$Li, both the molecular density and $Z$ become very small
towards the unitarity limit $\delta=0$. We will discuss the
physics of the system using two figures that summarize the main
results of our calculations.

\begin{figure}[htb]
\begin{center}
\includegraphics[width=8cm]{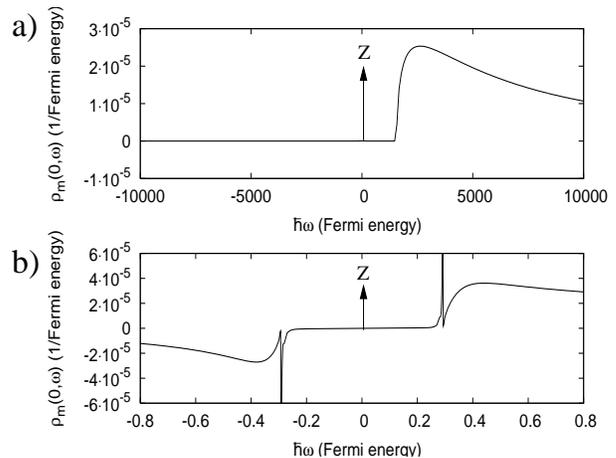}
\end{center}
\caption[a]{
Spectral functions of the bare $^6$Li$_2$ molecules with zero momentum a)
in the BEC limit at $571$ Gauss and b) in the BCS limit at $892$
Gauss of the crossover occurring near the broad Feshbach resonance
of $^6$Li at $834$ Gauss. The Fermi energy of the gas is $380$ nK.}
\label{BCSplaatje}
\end{figure}

In Fig.~\ref{BCSplaatje}, the spectral function for the bare
molecules is given for the BEC-side and the BCS-side of the
resonance. The top shows that the BEC-limit shows a structure that
resembles the two-body limit. The delta peak is the bare molecular
condensate that by definition exists at zero frequency. The
continuum of atomic scattering states causes a long tail on the
right side of the figure. The distance between these two features
is well approximated by the two-body bound state energy
$\epsilon_{\rm m}(\delta)$. The relevant energy scale in this
limit is much bigger than the Fermi energy, due to the extreme
broadness of the resonance, which is the reason that two-body
physics works so well.

In the bottom half of the same figure we present the spectral
density in the BCS-limit that shows completely different physics.
Again, the bare molecular condensate is at zero frequency, which
is now also the position of twice the Fermi surface, and the
two-quasiparticle continuum shows up as particle excitations on
the right, and hole excitations on the left. There are no
fermionic excitations within $2\Delta$ of twice the chemical
potential, as is expected from BCS theory. Left and right we see
also the peaks that are expected from RPA theory and physically
correspond to the long-lived modes associated with the
fluctuations in the magnitude of the BCS gap $\Delta$. A slight
curving at these ends anticipates these peaks, which is due to the
momentum dependence of the coupling constant $g$.

\begin{figure}[htb]
\begin{center}
\includegraphics[width=8cm]{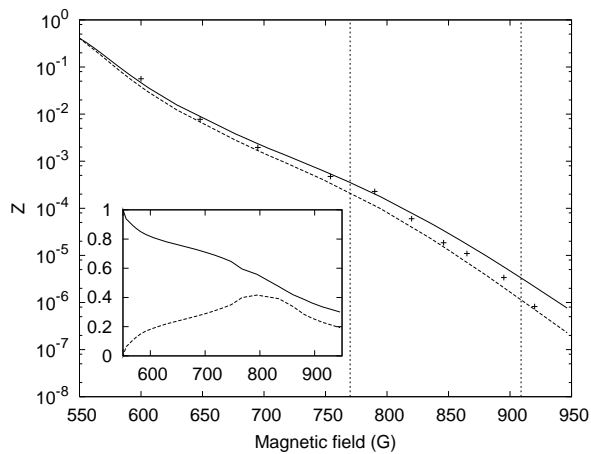}
\end{center}
\caption[a]{
The solid curve shows the probability $Z$ for $^6$Li. The data points are from the
experiment of Partridge {\it et al.} \cite{Rice}. The dashed curve
shows for reference the fraction $2Zn_{\rm mc}/n$ as a function of
magnetic field. The Fermi energy
$\hbar^2k_{\rm F}^2/2m$ of the gas is $380$ nK.  In the
inset the solid line shows the Bose-Einstein condensate fraction
of dressed molecules $2n_{\rm mc}/n$ and the dashed line the
contribution of the fluctuations to the total atomic density. The
vertical lines indicate the magnetic fields where $k_{\rm
F}|a_{\rm res}|=1$.}
\label{Z-plaatje}
\end{figure}

In Fig.~\ref{Z-plaatje} we then show with the solid line the
closed-channel fraction $Z$ as a function of magnetic field. This
is an improved version of the figure in Ref.~\cite{romans}, which
contained a numerical error in the determination of $Z$. A
comparison is made to the experimental data of Partridge {\it et
al.} \cite{Rice}. For reference, the bare molecular condensate
density fraction is also plotted. These quantities are not
identical and their ratio diverges as $\delta$ in the BCS-limit,
as we will show later on. The inverse of this ratio is equal to
the condensate fraction of dressed molecules, and it is plotted in
the inset. Fluctuations become very important on the positive side
of the Feshbach resonance, and their contribution to the total
density becomes of the same order as the condensate density.

\begin{figure}[htb]
\begin{center}
\includegraphics[width=8cm]{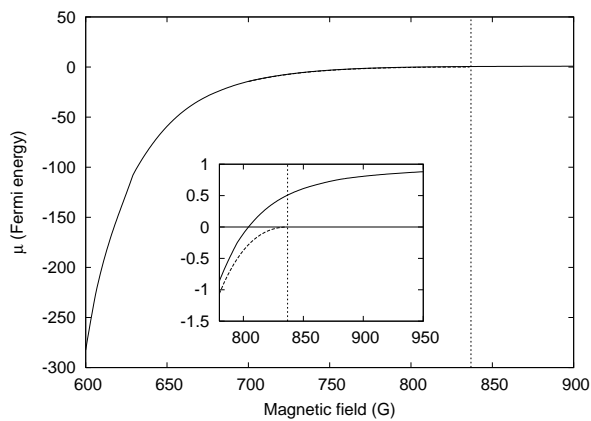}
\end{center}
\caption[a]{
Chemical potential as a function of magnetic field  for $^6$Li and for the same parameters as in Fig.~\ref{Z-plaatje}. At unitarity, we find $\mu  \simeq 0.51 \epsilon_F$. The dashed line is given by $\epsilon_{\rm m}(\delta)/2$, where $\epsilon_{\rm m}(\delta)$ is the energy of the dressed molecular state in two-body physics.}
\label{mu-plaatje}
\end{figure}
In Fig.~\ref{mu-plaatje} we see the behavior of the chemical potential as a function of the magnetic field. For low magnetic fields, the chemical potential is well approximated by half the binding energy of the two-body bound pair $\epsilon_{\rm m}(\delta)$. This corresponds well to our picture of having a pure Bose-Einstein condensate of dressed molecules in this limit. At high magnetic field, the chemical potential saturates and approaches the Fermi energy, as is expected for a fermionic gas. At zero detuning, we have that $\mu \simeq 0.51 \epsilon_{\rm F}$, which is commonly written as $(1+\beta)\epsilon_{\rm F}$. The result of $\beta  \simeq -0.49$ can then be compared to the Monte-Carlo result of $\beta \simeq -0.58$ \cite{astrak,montecarlo}.

\begin{figure}[htb]
\begin{center}
\includegraphics[width=8cm]{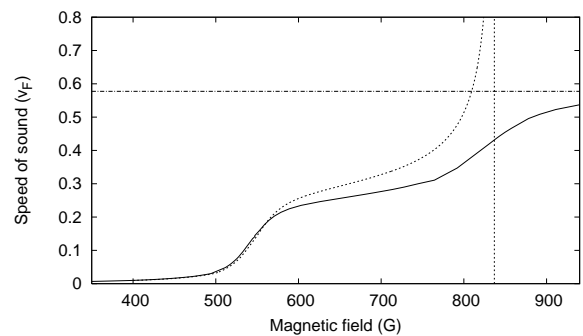}
\end{center}
\caption[a]{
Speed of sound in the case of $^6$Li and for the same parameters as in Fig.~\ref{Z-plaatje}. At unitarity, we find $c \simeq 0.43 v_{\rm F}$. The dashed line is the BEC-limit derived from Eq.~(\ref{generalcbec}) using Eqs.~(\ref{becZ}), (\ref{gapin2b}), and (\ref{hp120bec}). The dash-dotted line is the speed of sound $v_{\rm F}/\sqrt{3}$ of the Anderson-Bogoliubov mode.}
\label{sound-plaatje}
\end{figure}

The speed of sound, given in Fig.~\ref{sound-plaatje}, has an interesting
behavior, due to the interplay of the resonance, the background
scattering, and the dependence on the magnetic field of the
different parameters. Apart from the crossover near resonance, we
see around a magnetic field of 550 G, another crossover taking
place from the resonant-scattering dominated regime to the
background-scattering dominated regime. In the extreme BEC-limit,
the sound velocity goes to zero, as the gas approaches the limit of an ideal, noninteracting gas of molecules. On the right it approaches
the BCS-like limit of the Anderson-Bogoliubov mode. In the unitarity limit, we find $c \simeq 0.43 v_{\rm F}$. This sound velocity can be compared to previous work \cite{ghosh} where a unitarity value of $c \simeq 0.37 v_{\rm F}$ was found. One of the important reasons for this difference is the inclusion of background scattering effects in our approach. 

\section{Narrow resonances}
We now consider a much narrower resonance, in the absence of background scattering. From an
experimental viewpoint, this resonance with $\eta^2=\epsilon_{\rm
F}$, where $\epsilon_{\rm F}$ is the Fermi energy, is an extremely
narrow one, but from a theoretical point of view it is
intermediate.

\begin{figure}[htb]
\begin{center}
\includegraphics[width=8cm]{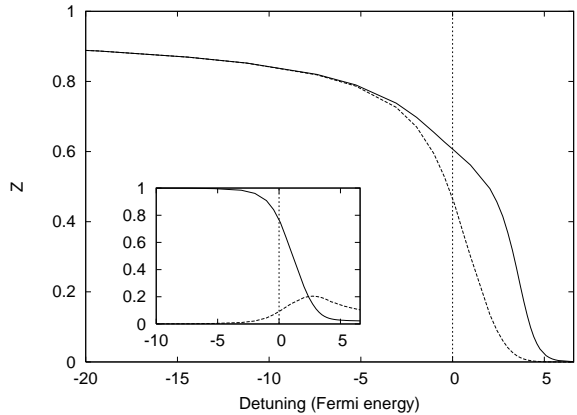}
\end{center}
\caption[a]{
The solid curve shows the probability $Z$ for a narrow resonance
in the absence of background interactions. The dashed curve shows
the fraction $2Zn_{\rm mc}/n$ as a function of magnetic field. In
the inset the solid line shows the Bose-Einstein condensate
fraction of dressed molecules $2n_{\rm mc}/n$ and the dashed line
the contribution of the fluctuations to the total atomic density.
The vertical line shows the position of the resonance.}
\label{narrowZ}
\end{figure}

\begin{figure}[htb]
\begin{center}
\includegraphics[width=8cm]{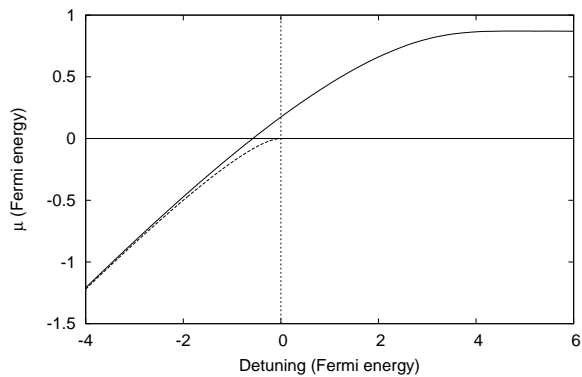}
\end{center}
\caption[a]{
Chemical potential as function of detuning for a narrow resonance ($\eta^2 = \epsilon_{\rm F}$). At unitarity, we find $\mu  \simeq 0.17 \epsilon_F$. The dashed line is given by $\epsilon_{\rm m}(\delta)/2$, where $\epsilon_{\rm m}(\delta)$ is the energy of the dressed molecular state in two-body physics.}
\label{munarrow-plaatje}
\end{figure}

Fig.~\ref{narrowZ} shows the closed-channel component $Z$,
and the bare molecular condensate density in this case. Again, the difference
between $Z$ and the bare condensate fraction is clearly visible.
What is interesting here, is that the crossover takes place at a somewhat higher detuning. At zero detuning we still have a large fraction of Bose-Einstein condensed dressed molecules, as they are stabilized by the Fermi-sea, which blocks the dissociation of the condensed molecules. This shift is also observable in Fig.~\ref{munarrow-plaatje}, where the chemical potential is plotted as a function of detuning. The chemical potential follows to a large extent the energy level of the bare molecule through the resonance, and only distinctly saturates at positive detuning. At resonance, we now find a lower value of $\mu \simeq 0.17 \epsilon_{\rm F}$. This shows that universality is lost for narrower resonances.   

\begin{figure}[htb]
\begin{center}
\includegraphics[width=8cm]{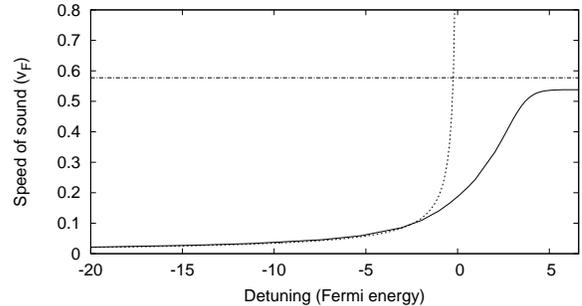}
\end{center}
\caption[a]{
Speed of sound in the case of a narrow resonance
($\eta^2=\epsilon_{\rm F})$. The dashed curve shows the BEC-limit
for a narrow resonance given in Eq.~(\ref{narrowsoundeq}). The
dash-dotted line is the speed of sound $v_{\rm F}/\sqrt{3}$ of the
Anderson-Bogoliubov mode.} \label{narrowsound}
\end{figure}

The speed of sound, shown in Fig.~\ref{narrowsound}, has a fairly
straightforward crossover behavior, mainly due to the absence of background
scattering. Again we retrieve the Anderson-Bogoliubov mode on the
right, and a vanishing speed of sound on the left.

\section{Various limits of the theory}
\begin{figure}[htb]
\begin{center}
\includegraphics[width=8cm]{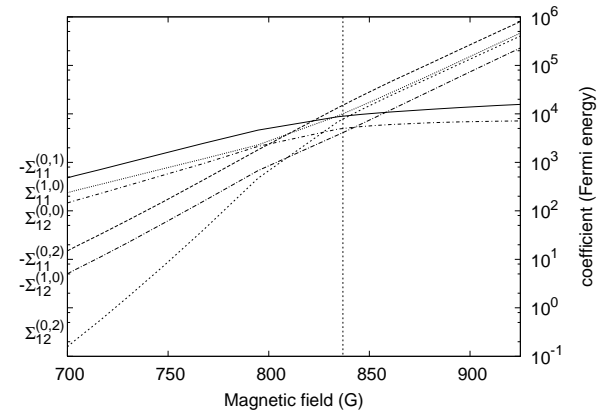}
\end{center}
\caption[a]{
Logarithmic plot of the coefficients in the Bogoliubov
Hamiltonian. The coefficients are defined in Eqs.~(\ref{defcoef}).
The term that dominates on the BEC-side is a first-order
derivative in energy ($\Sigma_{11}^{(0,1)}$), on the BCS-side it
is a second-order derivative in energy ($\Sigma_{11}^{(0,2)}$).
}\label{coefficients}
\end{figure}

In this section we study the three notable limits of the
crossover: the BEC-limit, the BCS-limit and the unitarity limit in more detail.
Focusing on the microscopic physics and the long-wavelength limit,
the crossover can be studied by looking at the behavior of the
self-energies for small energies. The various coefficients of
interest are shown in Fig.~\ref{coefficients}.

\subsection{BEC-limit}
In the BEC-limit, we expect to retrieve the results of two-body
physics. The reason for this is that, as a function of the
detuning, the chemical potential quickly becomes very negative,
eliminating the effects of the Fermi sea. The diagonal terms in
the many-body molecular self-energy reduce to the two-body
self-energies, and the off-diagonal self-energies become less
important. We define the BEC-limit by $|\Delta| \ll |\mu|$. To
leading order we derive from Eq.~(\ref{self-energy-elements})
\begin{align}
\hbar \Sigma_{11}(z,{\bf k})=
\frac{\eta\sqrt{-z+\epsilon({\bf k})/2-2\mu}}{1+\frac{\sqrt{m} |a_{\rm bg}|}{\hbar}\sqrt{-z+\epsilon({\bf k})/2-2\mu}}
\end{align}
in agreement with Eq.~(\ref{sigma2body}). Furthermore, we wish to
derive the expression for $Z$ given in Eq.~(\ref{Z2body}). This
follows directly from our general Eq.~(\ref{Z}) for $Z$ as
$\Sigma_{12}$ is small compared to the diagonal entry
$\Sigma_{11}$. Thus
\begin{align}
\label{becZ}
Z&=\frac{1}{1-\Sigma_{11}^{(0,1)}}=\left(1+\frac{\eta}{2\sqrt{-2\mu}(1+\frac{\sqrt{m} |a_{\rm bg}|}{\hbar}\sqrt{-2\mu})^2}\right)^{-1}
\end{align}
in agreement with Eq.~(\ref{Z2body}). In the BEC limit the
fluctuation terms in the equation of state can be neglected. This
then suggests that we have only a Bose-Einstein condensate of
dressed molecules, and nothing else. Our bare molecules are
contained  in the term $2\Delta^2/g^2$, the density of bare atoms
is given by the term ${\rm Tr}[G_{\rm a}\sigma_3]$. Indeed,
defining the density of Bose-Einstein condensed dressed molecules
as $n_{\rm mc}=n/2$, where $n$ is the total density, these terms
and the expression for $Z$ reduce to
\begin{align}
\label{gapin2b}
2\Delta^2/g^2&=2 Z n_{\rm mc},\\
{\rm Tr}[G_{\rm a}\sigma_3]&=2(1-Z)n_{\rm mc}.
\end{align}
This is illustrative for the fact that the dressed molecules can
be split in a molecular component with contribution $Z$, and an
atomic component with contribution $1-Z$.

The correct BEC limit for the speed of sound is slightly more
involved, as it is not a result of two-body physics. It explicitly
depends on the off-diagonal terms in the self-energy as we show
next. We can make a straightforward simplification by again using
$\Delta$ as our small parameter. Furthermore we use the fact that
$\Sigma_{11}^{(0,1)}=-2\Sigma_{11}^{(1,0)}$ in this limit. The
general expression for the speed of sound is then
\begin{align}
\label{generalcbec}
c&=\left[\frac{\hbar\Sigma_{12}^{(0)}}{2m\left(1-\Sigma_{11}^{(0,1)}\right)}\right]^{1/2}.
\end{align}
Depending on the background scattering length, the width of the
resonance, and the density, the system can enter different
regimes. Expanding in the parameter $\Delta/|\mu|$, the
off-diagonal term reduces to
\begin{align}
\label{hp120bec}
\hbar\Sigma_{12}^{(0)}=\frac{\Delta^2\eta}{16(-\mu)^{3/2}(1 + 2\mu/E_{\rm bg})^4}\Bigg(32\left(\frac{-\mu}{E_{\rm bg}}\right)^{5/2} -\nonumber\\
 60 \sqrt{2}\left(\frac{\mu}{E_{\rm bg}}\right)^2 +80\left(\frac{-\mu}{E_{\rm bg}}\right)^{3/2} + 20 \sqrt{2}\frac{\mu}{E_{\rm bg}} + \sqrt{2}\Bigg),
\end{align}
where $E_{\rm bg}=\hbar^2/m a_{\rm bg}^2$. The gap is
fundamentally a many-body property, but it can be related to the
two-body $Z$ and the density through Eq.~(\ref{gapin2b}). One
possible regime is $|\mu| \ll E_{\rm bg}$, where the background
scattering is small, while being sufficiently close to resonance
such that $|\delta| \ll \eta^2$, yet sufficiently far to be in the
BEC-limit. In that regime the resonant scattering determines the
speed of sound, and we find the simple expression
\begin{align}
c=\hbar\sqrt{2 \pi a_{\rm res} n_{\rm mc}}/m,
\end{align}
where $a_{\rm res}=\hbar \eta / \delta \sqrt{m}=\hbar \eta / \Delta\mu (B-B_0) \sqrt{m}$ is the resonant scattering length, and $n_{\rm mc}$ is the Bose-Einstein condensate density of dressed molecules. This can be cast in the more familiar result
\begin{align}
\label{familiar}
c=\hbar\sqrt{4 \pi a_{\rm m} n_{\rm mc}}/m_{\rm m},
\end{align}
using the relevant parameters for molecules $a_{\rm m}=2 a_{\rm
res}$ \cite{demelo} and $m_{\rm m} = 2m$. This is precisely the
Bogoliubov speed of sound \cite{bogoliubovc}. Note that the
Bose-Einstein condensate density of dressed molecules shows up in
the expression for the speed of sound, further strengthening the
case that the dressed molecules are the relevant physical entities
in this system. In the case of $^6$Li, at densities that are
typical for experiment, this regime is however not well separated from other regimes,
due to the extremely large background scattering.

\begin{figure}[htb]
\begin{center}
\includegraphics[width=8cm]{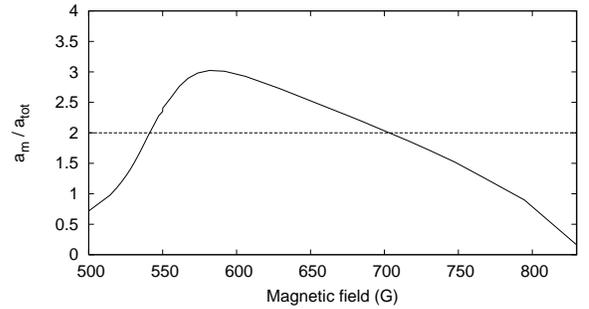}
\end{center}
\caption[a]{
Ratio of the molecular and total atomic scattering length as a
function of magnetic field.} \label{singlea}
\end{figure}

Before we move on to the background-scattering dominated regime, we compare our results to single-channel theory. In this case there is only a single
parameter, the total atomic scattering length $a= a_{\rm
res}+a_{\rm bg}$. Assuming the validity of Eq.~(\ref{familiar}),
the molecular scattering length can be derived from the speed of
sound, and expressed in terms of the total atomic scattering
length $a$. This is shown in Fig.~\ref{singlea}. The expected
one-loop single-channel result $a_m=2a$ is given by the dashed
line and can basically never be considered as a good approximation
to the one-loop result obtained from our two-channel theory. We
expect the same to be true of the exact single-channel result.

When the background scattering gives a significant contribution,
and $|\mu| \simeq E_{\rm bg}$, which for $^6$Li at the densities
of interest happens roughly around a magnetic field of 650 G, it
has a strong effect on the speed of sound. Taking it into account
using Eq.~(\ref{hp120bec}), and relating the speed of sound to the
two-body $Z$ as described above, we get the result that is drawn
by the dashed line in Fig.~\ref{sound-plaatje}. We can conclude that the
speed of sound is not so well approximated with two-body physics
as the closed-channel component $Z$. This is maybe to be expected
as the speed of sound is really a many-body property of the gas.

In the extreme BEC-limit, the speed of sound vanishes as
\begin{align}
c&=\left(\frac{\eta g^2 n E_{\rm bg}^{3/2}}{2 [-\epsilon_{\rm m}(\delta   )]^3}\right)^{1/2},
\end{align}
where $\epsilon_{\rm m}(\delta)$ is the energy of the exact bound
molecular state taken from Eq.~(\ref{be}). In this way the theory
is completely reduced to two-body physics, where there indeed
exists no sound. Of course, the vanishing of the speed of sound in
the extreme BEC limit is a result of the fact that we neglected
the background scattering between molecules. If this is
effectively repulsive, it will ultimately stabilize the molecular
condensate and result in a nonzero speed of sound. Finally, in
theory we can also be far away from resonance, and in a regime
where the effect of the atomic background scattering is small. In
this case the speed of sound reduces to
\begin{align}
\label{narrowsoundeq}
c&=
\left(\frac{\eta g^2 n}{8 (-\delta )^{3/2}}\right)^{1/2}.
\end{align}
This solution is shown by the dashed line in Fig.~\ref{narrowsound}.

\subsection{BCS-limit}
For large positive detuning, the system strongly resembles that of
a normal BCS superfluid. The density of bare molecules becomes
exponentially small, and the properties of the system depend
mainly on the gap parameter $\Delta$, which is also decreasing
exponentially. Moreover, the chemical potential tends to the Fermi
energy.

Again, we can study the system by looking at the action for long
wavelengths. We find explicitly,
\begin{equation}
\begin{split}
\hbar\Sigma_{11}^{(0)}&\simeq\phantom{-}\frac{2\sqrt{2}}{\pi}\eta\frac{\sqrt{\epsilon_{\rm F}}}{1+2\epsilon_{\rm F}/E_{\rm bg}}\log(\Delta/\epsilon_{\rm F}),\\
\hbar\Sigma_{12}^{(0)}&\simeq\phantom{-}\frac{\sqrt{2}}{\pi}\eta\frac{\sqrt{\epsilon_{\rm F}}}{1+2\epsilon_{\rm F}/E_{\rm bg}},\\
\hbar\Sigma_{11}^{(0,1)}&\simeq\phantom{-}\frac{2\sqrt{2}}{\pi}\frac{\eta}{\sqrt{\epsilon_{\rm F}}}\frac{1-2\epsilon_{\rm F}/E_{\rm bg}}{4(1+2\epsilon_{\rm F}/E_{\rm bg})}\log(\Delta/\epsilon_{\rm F}),\\
\hbar\Sigma_{11}^{(0,2)}&\simeq-2\sqrt{2\epsilon_{\rm F}}\eta/3\pi\Delta^2\\
\hbar\Sigma_{12}^{(0,2)}&\simeq\phantom{-}\sqrt{2\epsilon_{\rm F}}\eta/3\pi\Delta^2\\
\hbar\Sigma_{11}^{(1,0)}&\simeq\phantom{-}4\sqrt{2}\epsilon_{\rm F}^{3/2}\eta/9\pi\Delta^2\\
\hbar\Sigma_{12}^{(1,0)}&\simeq-2\sqrt{2}\epsilon_{\rm F}^{3/2}\eta/9\pi\Delta^2,
\end{split}
\end{equation}
where $\epsilon_{\rm F}$ is the Fermi energy. We see that the
coefficients $\Sigma_{11}^{(0,2)}$, $\Sigma_{12}^{(0,2)}$,
$\Sigma_{11}^{(1,0)}$ and $\Sigma_{12}^{(1,0)}$ diverge most
strongly for small $\Delta$. We can use Eq.~(\ref{sound}) to find
the speed of sound in this limit and we obtain
\begin{align}
\label{andersound}
c=\frac{v_{\rm F}}{\sqrt{3}},
\end{align}
where $v_{\rm F}$ is the Fermi velocity $\hbar k_{\rm F}/m$. This
coincides with the speed of sound of the Anderson-Bogoliubov mode
\cite{anderson,bogoliubov,galitskii}. A more illustrative way of
deriving this result is to determine the eigenvalues of the
propagator in the long-wavelength limit. The leading terms are the
ones mentioned above, leading to
\begin{align}
S_{\rm m}&[\psi_{\rm m}^*,\psi_{\rm m}]
\simeq\frac{\sqrt{2 \epsilon_{\rm F}}\eta}{36\pi\Delta^2}
\sum_{{\bf k},n}
\left[
\psi_{\rm m}^*,\psi_{\rm m}
\right]\\
&\times\left[\begin{array}{cc}
\phantom{-}6 (\hbar\omega)^2 + 8 \epsilon_{\bf k} \epsilon_{\rm F}&
-3 (\hbar\omega)^2 -4\epsilon_{\bf k} \epsilon_{\rm F}\\
-3 (\hbar\omega)^2 -4\epsilon_{\bf k} \epsilon_{\rm F} &
\phantom{-}6 (\hbar\omega)^2 + 8 \epsilon_{\bf k} \epsilon_{\rm F}\\
\end{array}\right]
\left[\begin{array}{c}
\psi_{\rm m}\\
\psi^*_{\rm m}\\
\end{array}\right].\nonumber
\end{align}
This action can be diagonalized by introducing amplitude and phase
fluctuations. Keeping in mind that $\psi_{\rm m}$ are the
fluctuations of the molecular field, we define $\psi_{\rm
m}=\rho/g+i \Delta \theta/g$ and $\psi^*_{\rm m}=\rho/g-i \Delta
\theta/g$, to obtain
\begin{align}
S_{\rm m}[\rho, \theta]&\simeq\frac{\sqrt{2 \epsilon_{\rm F}}\eta}{18\pi  g^2 \Delta^2}
\sum_{{\bf k},n}
\left[
\rho^*,\Delta\theta^*\right]\times\nonumber\\
&\left[\begin{array}{cc}
3 (\hbar\omega)^2 + 4 \epsilon({\bf k}) \epsilon_{\rm F}&0\\
0&9 (\hbar\omega)^2 + 12 \epsilon({\bf k}) \epsilon_{\rm F}\\
\end{array}\right]
\left[\begin{array}{c}
\rho \\
\Delta \theta \\
\end{array}\right]\nonumber\\
&=\frac{N(0)}{36}
\sum_{{\bf k},n}
\left(
\frac{|\rho({\bf k},i\omega_n)|^2}{\Delta^2}+3|\theta({\bf k},i\omega_n)|^2\right)
\nonumber\\
&\times
\left(3 (\hbar\omega)^2 + 4 \epsilon({\bf k}) \epsilon_{\rm F}\right),
\end{align}
in agreement with Ref.~\cite{stoof1993}. Here $N(0)$ is the
density of states of a single spin state at the Fermi surface. We
expect that the non-leading terms in $\Delta$ lead to a mass term
of order $\Delta^2$ for the amplitude fluctuations. The result in
Eq.~(\ref{andersound}) can be immediately derived from the phase
fluctuation part.

The closed-channel component in the BCS-limit is
\begin{align}
Z=\frac{\pi (1-2 \epsilon_{\rm F}/E_{\rm bg})}{2
   \sqrt{2} \eta  \epsilon_{\rm F}^{3/2}}  \Delta ^2 \log
\left(\epsilon_{\rm F}/\Delta \right) .
\end{align}
It is important to notice that in this BCS limit, $Z$ and the
density contribution of the Bose-Einstein condensate of bare
molecules $2 \Delta^2/g^2$ have an entirely different behavior as
a function of detuning. From the gap equation it can be derived
that the gap vanishes exponentially,
\begin{align}
\Delta \propto \epsilon_{\rm F} \exp\left\{-\frac{\pi  (1+ 2
\epsilon_{\rm F}/E_{\rm bg})}{2 \eta  \sqrt{2 \epsilon_{\rm F}
}}\delta
 \right\},
\end{align}
thus the ratio between $Z$ and $\Delta^2$ diverges linearly with
the detuning. This shows that these are different entities with
different physical meanings.

\subsection{Unitarity limit}
For $\delta=0$ we can make a rough estimate of $Z$, that gives a
simple physical reason for why $Z$ is small for broad Feshbach
resonances. As can be seen in Fig.~\ref{BCSplaatje}, the most
notable result of the many-body theory is the gap around the Fermi
surface. In contrast to two-body physics, where $Z$ goes to zero
at resonance, the many-body physics stabilizes the pairs, much
like Cooper-pairs. For broad resonances, the spectral function has
a long tail at high energies, which is not affected by many-body
physics because it lies far above the Fermi energy. Using the fact
that the spectral function is normalized to one, we can thus
approximate the spectral weight in the delta function by
integrating the two-body spectral function over an energy interval
of twice the gap.
\begin{align}
\label{unitZ}
Z\simeq\frac{1}{\pi}\int_0^{2\Delta} \frac{\sqrt{z} \eta }{z^2+\eta ^2 z} dz
\simeq \frac{2 \sqrt{2\Delta }}{\pi \eta }
\end{align}
From our calculations, the gap at unitarity is equal to $0.46
\epsilon_{\rm F}$, which is quite close to the Monte-Carlo result
of $\Delta \simeq 0.50 \epsilon_{\rm F}$ \cite{montecarlo}.
Substituting this in Eq.~(\ref{unitZ}) results in an approximate
value for $Z$ of $3.5 \cdot 10^{-5}$, which is only about 20\% lower
than the result of our Bogoliubov theory.

\section{Conclusions}
The BEC-BCS crossover near a Feshbach resonance is very rich in
physics. We have presented a theory that rigorously incorporates
the two-body Feshbach physics and, due to the inclusion of
fluctuations, allows for an accurate study of the full crossover
problem, both for broad and narrow resonances. We calculated the
probability $Z$, that is crucial in understanding the microscopic
physics of the crossover, and studied the molecular spectral
function and the spectrum of collective modes throughout the
BEC-BCS crossover. Where possible, we have compared our theory with experiment and have found a satisfying agreement, even in the unitarity limit where the interactions are very strong.

\section*{Acknowledgments}
This work was supported by the Stichting voor
Fundamenteel Onderzoek der Materie
(FOM), which is supported by the Nederlandse Organisatie voor Wetenschappelijk
Onderzoek (NWO).


\end{document}